\documentclass[journal=rsm]{CUP-JNL-DTM}%

\usepackage{graphicx}
\usepackage{multicol,multirow}
\usepackage{amsmath,amssymb,amsfonts}
\usepackage{mathrsfs}
\usepackage{amsthm}
\usepackage{rotating}
\usepackage{appendix}
\usepackage{ifpdf}
\usepackage[T1]{fontenc}
\usepackage{newtxtext}
\usepackage{newtxmath}
\usepackage{textcomp}
\usepackage{xcolor}
\usepackage{lipsum}
\usepackage[colorlinks,allcolors=blue]{hyperref}

\def\A{\mathbf A}
\def\bC{\mathbf C}
\def\D{\mathbf D}
\def\I{\mathbf I}

\def\L{\mathbf L}
\def\M{\mathbf M}
\def\T{\mathbf T}
\def\X{\mathbf X}
\def\W{\mathbf W}
\def\H{\mathbf H}
\def\d{\mathbf d}
\def\e{\mathbf e}
\def\p{\mathbf p}

\def\y{\mathbf y}
\def\0{\mathbf 0}
\def\1{\mathbf 1}

\theoremstyle{definition}

\jname{Research Synthesis Methods}
\articletype{RESEARCH ARTICLE}
\jyear{2026}

\begin{document}

\begin{Frontmatter}

\title[Article Title]{Network meta-analysis and diffusion}

\author[1]{Gerta R\"ucker}
\author[2]{Annabel L. Davies}
\author[1]{Guido Schwarzer}

\authormark{R\"ucker \textit{et al}.}

\address[1]{\orgdiv{Institute of Medical Biometry and Statistics}, \orgname{Faculty of Medicine and Medical Center - University of Freiburg}, \orgaddress{\city{Freiburg}, \postcode{79104}, \country{Germany}}. \email{gertaruecker@aol.com}}

\address[2]{\orgdiv{Department of Population Health Sciences, Bristol Medical School}, \orgname{University of Bristol}, \orgaddress{\city{Bristol}, \country{UK}}}

\authormark{R\"ucker et al.}



\abstract{We show that the covariance matrix of the treatment effect estimates in a network meta-analysis can be obtained without matrix inversion using a geometric series of diffusion matrices. This property extends to the hat matrix and provides a connection between parameter estimation in regression analysis and random walks on the network graph. We also provide a number of visualization tools implemented in R.}

\end{Frontmatter}


\section*{Highlights}

\subsection*{What is already known}
Network meta-analysis is a powerful tool in evidence-based medicine.

\subsection*{What is new}
The covariance matrix of the treatment effect estimates in a network meta-analysis can be obtained using a geometric series of diffusion matrices.

\subsection*{Potential impact for Research Synthesis Methods readers}
Our result provides new insights into the connection between network meta-analysis, diffusion and random walks on networks.


\section{Introduction} \label{intro}

Network meta-analysis (NMA) is a powerful tool in evidence-based medicine. Users may choose between frequentist and Bayesian methods, and software for both is  widely available. Both contrast-based and arm-based methods have been implemented.\cite{Ades:Welt:Dias:Phil:Cald:twen:2024}

Contrast-based frequentist methods for NMA can be viewed from two perspectives, the regression perspective and the graph-theoretical perspective. Although formally equivalent and leading to exactly the same results, both perspectives provide different insights.\cite{Rucke:Schwa:2014} In this paper, we show that two characteristic objects of weighted least squares regression, the variance-covariance matrix and the hat matrix, can be obtained using diffusion, a concept from graph theory related to random walks.

The article is structured as follows. After introducing three examples in Section \ref{examples}, we present our concepts in Section \ref{methods}. In Section \ref{interpretation} we suggest physical models to understand the connection between the different concepts. We apply the methods to the examples (Section \ref{results}), using R package \textbf{netmeta} \cite{Baldu:Rcke:Nikol:Papak:Salan:Efthi:Schwa:netmeta:2023,netmeta:2026} and a number of additional R functions for visualization, described in the Appendix and found in the online supporting information. The paper ends with a discussion (Section \ref{discussion}). 

\section{Examples} \label{examples}

We use a small toy example to illustrate the methods and apply them to two real-data NMAs.

\subsection{Example 1} \label{example1}

\begin{figure}[h!tb]
  \caption{Example network graphs. Left panel: A fictitious network with five treatments and seven studies. Mid panel: Network graph of the Dong 2013 data. ICS, inhaled corticosteroid; LABA, long-acting $\beta2$ agonist; TIO-HH, tiotropium dry powder delivered via HandiHaler; TIO-SMI, tiotropium solution delivered via Resipmat Soft Mist Inhaler. Right panel: Network graph of the Jalota 2011 data. Ante, Antecubital vein; Hand, Hand vein; Lido-prop, Lidocaine-propofol admixture; Lido-pre, Lidocaine pretreatment; Keta-pre, pretreatment with ketamine; NSAIDS-pre, pretreatment with non-steroidal anti-inflammatory drugs; Opioid-pre, Pretreatment with opioids.}
  \label{netgraphs}
  \begin{center}
    \includegraphics[width=\textwidth]{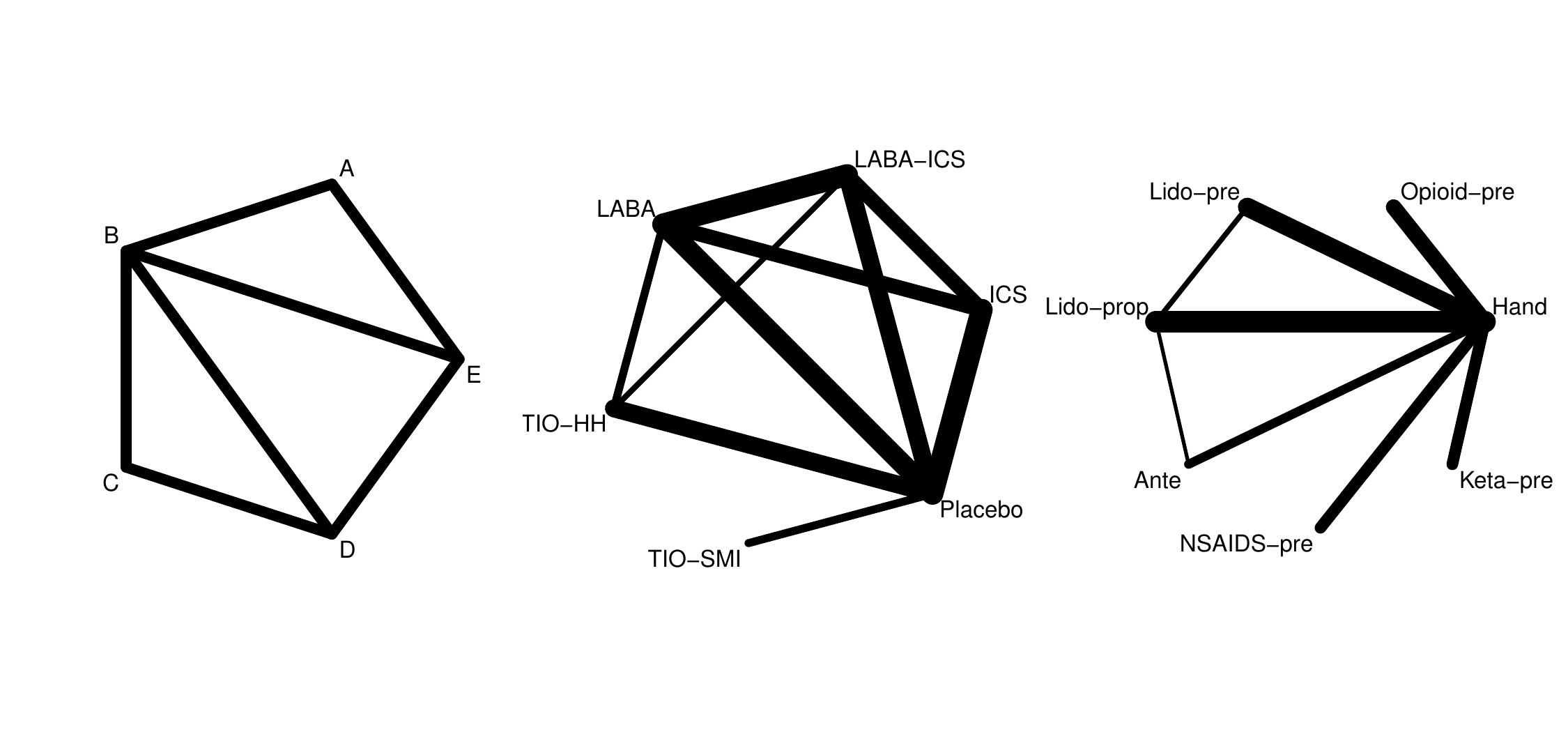}
  \end{center}
\end{figure}

The toy example with five treatments and seven two-arm studies is shown in the left panel of Figure~\ref{netgraphs}. All variances are assumed to be 1. The observed treatment effects for the comparisons A:B, A:E, B:C, B:D, B:E, C:D, D:E are also assumed to be equal to 1.

\subsection{Example 2} \label{example2}

The second example is an NMA of 41 trials (including seven four-arm studies and three three-arm studies, making 82 pairwise comparisons in total) investigating the safety of six inhaled medications, including placebo, in patients with chronic obstructive pulmonary disease. The outcome was mortality, measured as binary, with the odds ratio as the effect measure.\cite{Dong:Lin:Shau:Wu:Chang:Lai:2013} The data set is available from the R packages \textbf{netmeta} and \textbf{metabook}.\cite{netmeta:2026,metabook:2026} The network graph is shown in the middle panel of Figure~\ref{netgraphs}.

\subsection{Example 3} \label{example3}

The third example, shown in the right panel of Figure~\ref{netgraphs}, is a network of 102 trials (including 3 three-arm trials) comparing 7 drug and non-drug interventions, including placebo, to prevent pain on propofol injection.\cite{Jalota:2011} The binary outcome was pain, with relative risk (RR) as a measure of effect. The data set is available from R package \textbf{metabook}.\cite{metabook:2026}

\section{Methods} \label{methods}

Let $\X$ be the design matrix of a network meta-analysis (NMA) where the rows represent studies or comparisons and the columns represent the $n$ treatments, represented as nodes in the network graph. Each row of $\X$ contains $1$ in the column corresponding to one of the treatments in the row, and $-1$ in the column corresponding to the other treatment, and zeros elsewhere. Let $\W$ be the diagonal matrix of inverse variance weights, after adjusting weights for multi-arm trials, where necessary.\cite{Rcke:netw:2012} 

\paragraph{Example 1} For our toy example, the weight matrix is equal to $\W = \I$, the $7 \times 7$ identity matrix, and the design matrix is
\begin{align*}
\X =
\begin{pmatrix} 
1 & -1 & \phantom{-}0 & \phantom{-}0 & \phantom{-}0 \\ 
1 & \phantom{-}0 & \phantom{-}0 & \phantom{-}0 & -1 \\ 
0 & \phantom{-}1 & -1 & \phantom{-}0 & \phantom{-}0 \\ 
0 & \phantom{-}1 & \phantom{-}0 & -1 & \phantom{-}0 \\ 
0 & \phantom{-}1 & \phantom{-}0 & \phantom{-}0 & -1 \\ 
0 & \phantom{-}0 & \phantom{-}1 & -1 & \phantom{-}0 \\ 
0 & \phantom{-}0 & \phantom{-}0 & \phantom{-}1 & -1    
\end{pmatrix}
\end{align*}
where the columns correspond to the five treatments (nodes) A, B, C, D, E and the rows to the 7 comparisons listed in Subsection \ref{example1} and shown in Figure~\ref{netgraphs}, left panel. Alternatively, we may use the full design matrix including the non-existent edges AC, AD and CE and for $\W$ the $10 \times 10$ identity matrix with zeros at the diagonal positions of the non-existent edges. For real data, the study-based design matrix will be used, such that the rows correspond to studies and studies that compare the same treatments contribute identical rows.

We now describe two $n \times n$ matrices, the covariance matrix and the hat matrix of an NMA. The \emph{covariance matrix} of the NMA-based treatment effect estimates is obtained as
$$\bC = \X \ (\X ^\top \W \X)^+ \X ^\top = \X \ \L^+ \X^\top $$
where $\L = \X ^\top \W \X$ is the Laplacian matrix and $^+$ denotes the Moore-Penrose pseudoinverse.\cite{Rcke:netw:2012,Rucke:Schwa:2014,Albe:regr:1972}
The \emph{hat matrix} $\H$ can be written as 
$$\H = \bC \W = \X \ (\X ^\top \W \X)^+ \X ^\top \W = \X \ \L^+ \X^\top \W.$$
If $\W = \I$, the identity, we have $\H = \bC$.

Each diagonal element of $\L$ corresponds to a treatment (node) in the network and is equal to the sum of the weights in the edges connected to that node, known as the weighted degree of the node. 
Therefore, we use the diagonal of $\L$ to define a vector, $\d = \mbox{diag}(\L)$, and a diagonal matrix, $\D$, of weighted degrees. The diagonal elements of $\D$ are equal to $\d$ and the off-diagonal elements are equal to 0.
The weighted adjacency matrix of the network is then given by
$$\A = \D - \L.$$
For the unweighted case ($\W = \I$), $\A$ is the usual adjacency matrix. 

Now we define a transition matrix, also called a diffusion matrix, by
\begin{align}\label{eq:T}
\T = \A \D^{-1} = (\D - \L) \D^{-1} = \I - \L \D^{-1}.
\end{align}
This matrix describes a random diffusion process on the network.
The concept of diffusion, related to that of random walks, plays a role in graph theory and its applications, particularly physics,\cite{Spie:spec:2012,Spiel:2025} and we applied it to NMA in an earlier paper.\cite{Davie:Papak:Nikol:2022}
Consider a random walker moving between nodes of the network; each element $\T_{ij}$ of the diffusion matrix describes the probability of a walker who is currently at node $j$, transitioning to node $i$ in the next step.
According to Equation (\ref{eq:T}), this transition probability is proportional to the weight associated with the edge connecting nodes $i$ and $j$ (normalized by the weighted degree of node $j$).
Therefore, the walker is more likely to move along edges with larger weight.

\paragraph{Example 1} For the toy example, we have
\begin{align*}
\D = 
\begin{pmatrix} 
2  &  0  &  0  &  0  &  0 \\
0  &  4  &  0  &  0  &  0 \\
0  &  0  &  2  &  0  &  0 \\
0  &  0  &  0  &  3  &  0 \\
0  &  0  &  0  &  0  &  3 \\
\end{pmatrix}, \quad
\L =
\begin{pmatrix} 
\phantom{-}2  &  -1  &  \phantom{-}0  &  \phantom{-}0  &  -1 \\
-1  &  \phantom{-}4  &  -1  &  -1 &  -1 \\
\phantom{-}0  &  -1  &  \phantom{-}2  &  -1  &  \phantom{-}0 \\
\phantom{-}0  &  -1  &  -1  &  \phantom{-}3  &  -1 \\
-1  &  -1  &  \phantom{-}0  &  -1  &  \phantom{-}3 \\
\end{pmatrix}, \quad
\T = 
\begin{pmatrix} 
0  &  \frac{1}{4}   &  0  &  0  &  \frac{1}{3} \\
\frac{1}{2}  &  0  &  \frac{1}{2}   &  \frac{1}{3}   &  \frac{1}{3} \\
0  &  \frac{1}{4}   &  0  &  \frac{1}{3}   &  0 \\
0  &  \frac{1}{4}   &  \frac{1}{2}   &  0  &  \frac{1}{3} \\
\frac{1}{2}   &  \frac{1}{4}   &  0  &  \frac{1}{3}   &  0 \\
\end{pmatrix}.
\end{align*}
The columns of $\T$ sum up to 1 and describe for each column (node) a multinomial distribution with probabilities for moving from the given node to each of its neighbors. For an unweighted network, non-zero probabilities within the same column are equal; for a general NMA they become different. These probabilities provide the basis for the infinite diffusion process we describe in the next section.

\subsection{A geometric sequence of diffusion matrices} \label{sequence}

For the following, we first assume that the network graph contains at least one loop of odd length, in other words, it is non-bipartite. We consider the geometric sequence of the powers of the diffusion matrix, $(\T^i)_{i = 0, 1, 2, \dots}$. Under the assumption that this sequence is converging, its limit is the matrix
$$\T^{\infty} = \d_0 \ \1^\top$$
where $\1$ is the vector of ones of length $n$ and $\d_0$ is the standardized vector of degree proportions, $\d_0 = \d/||\d||$ where $||\d||$ is the sum of the weighted degrees over all nodes. $\T^{\infty}$ has $n$ equal columns that are all equal to $\d_0$.\cite{Spiel:2025} 

If $\p = (p_1, \dots, p_n)$ is a starting vector of probabilities $0 \le p_i \le 1, i = 1, \dots, n$ and $\sum p_i = 1$, we have under the assumption of convergence
\begin{align}\label{stationarydist}
\T^{\infty} \p = \d_0 \ \1^\top \p =  \d_0 \sum p_i =  \d_0,
\end{align}
justifying the notion of $\T$ as a diffusion matrix: starting from any distribution of a constant mass over the nodes, iterating $\T$ leads to a stable limit distribution of the mass such that each node obtains a mass proportional to its weighted degree.

\begin{figure}[h!tb]
  \caption{Diffusion process for Example 1, starting in node A.}
  \label{example1-diffusion}
  \begin{center}
    \includegraphics[width=13cm]{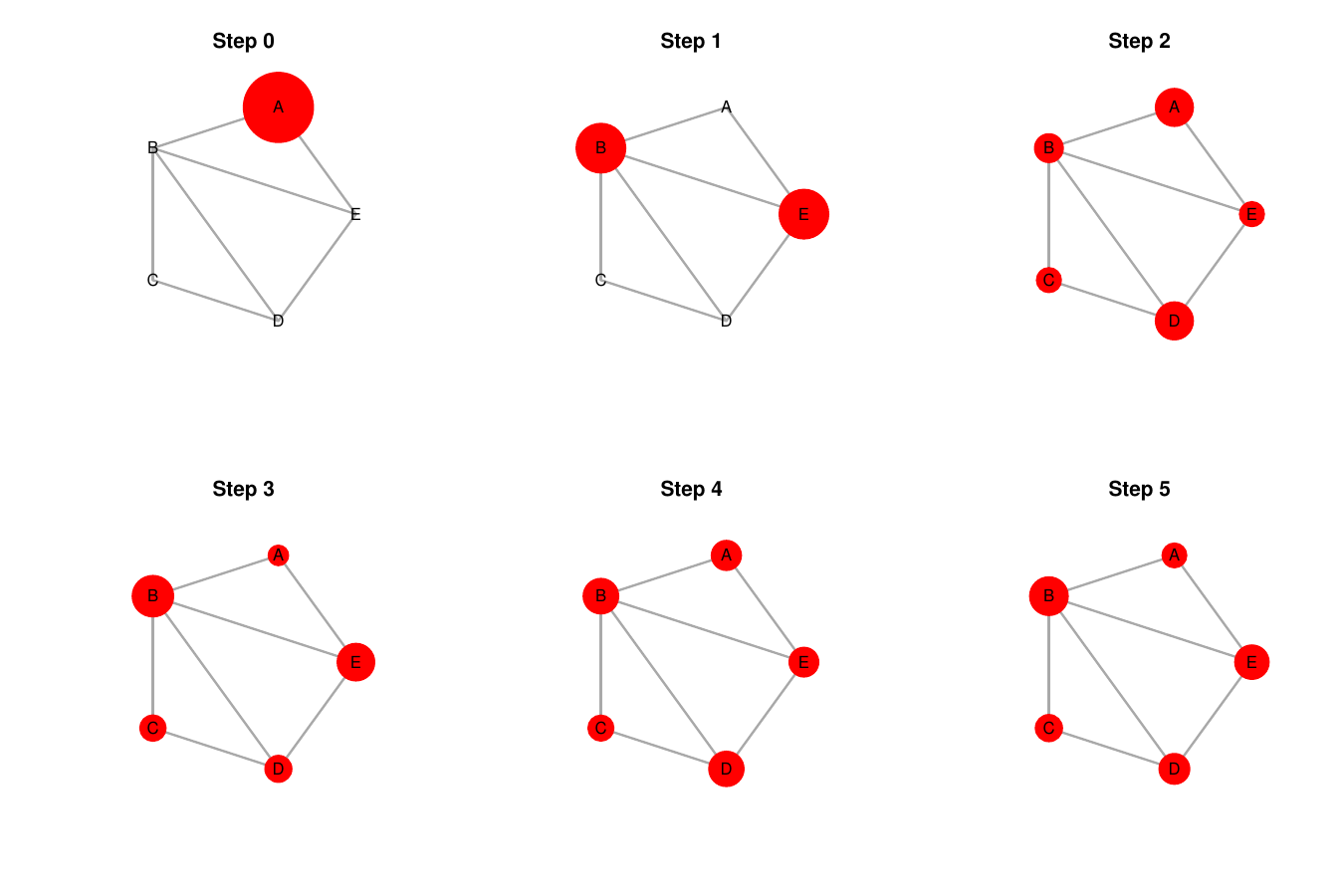}
  \end{center}
\end{figure}

\paragraph{Example 1}
We apply the diffusion process to the toy example (Figure~\ref{netgraphs}, left panel). Figure~\ref{example1-diffusion} shows the first six steps of the diffusion process when starting in node A (step 1). 
In the first step, the mass is equally divided between nodes B and E. 
In the following steps, the mass is dispersed across all available nodes. 
The limit distribution is reached with a precision of seven decimal places after 34 steps (see R code in the supporting information).
It is given by
\begin{align*}
\T^\infty = \frac{1}{14}
\begin{pmatrix} 
2 &  2  &  2  &  2  &  2 \\
4 &  4  &  4  &  4  &  4 \\
2 &  2  &  2  &  2  &  2 \\
3 &  3  &  3  &  3  &  3 \\
3 &  3  &  3  &  3  &  3 \\
\end{pmatrix}.
\end{align*}

We note that we did not prove the convergence of the sequence $(\T^i)_{i = 0, 1, 2, \dots}$ and the existence of $\T^\infty$. In fact, for bipartite network graphs this sequence does not converge but oscillates. We will treat this case in the next section \ref{bipartite}.

\subsection{Bipartite networks} \label{bipartite}

A bipartite graph is a graph that does not contain odd-numbered loops (cycles). As a result, the nodes can be split into two sets such that nodes from each set are exclusively connected to nodes from the other set and vice versa. Examples are networks without any loops, also called trees, such as star-shaped networks, or networks containing only even-numbered loops, all of which are frequently occurring in NMA.

For bipartite networks, the geometric sequence of diffusion matrices, as defined above, does not converge. Instead, the diffusion matrices oscillate between the two sets of nodes and thus do not lead to a stable limit distribution. Oscillation is a well-known feature of bipartite graphs.\cite{Bonchev:Kier:1992,Rcke:Rcke:math:1994} Fortunately, there is a pragmatic solution: We can define a so-called \emph{lazy walk}, which means that in each step half of the mass remains in the current node and the other half continues to diffuse according to the edge weights. Following Spielman,\cite{Spie:spec:2012,Spiel:2025} the transition matrix is then
$$ \tilde \T = (\T + \I)/2$$
and the limit of the sequence $\tilde \T^i$ again becomes
$$\T^{\infty} = \d_0 \ \1^\top.$$
As this is the more general version of diffusion, we use a lazy walk in most of the R functions described in Section \ref{R-code}. By contrast to the lazy walk, we refer to the walk defined by $\T$ in equation \eqref{eq:T} as the \emph{simple walk}.

\subsection{A geometric series of diffusion matrices} \label{series}

If the sequence $\tilde \T^i \rightarrow \T^{\infty} \ne \0$ converges, the geometric series of the partial sums $\sum_{i=0}^{N} \tilde \T^i \ (i = 0, 1, 2, \dots, N)$ diverges. However, we can state the following theorem.

\paragraph{Theorem} For both bipartite and non-bipartite networks, the covariance matrix can be written using a geometric series of diffusion matrices:
\begin{align} \label{covariance}
\bC = \frac{1}{2} \X \D^{-1} \sum_{i=0}^{\infty} \left( \tilde \T^i  \X^\top \right). 
\end{align}
For a proof of equation \eqref{covariance}, see Appendix \ref{proof}. 

\paragraph{Corollary} For the hat matrix, we have
\begin{align} \label{hatmatrix}
\H = \frac{1}{2} \X \D^{-1} \sum_{i=0}^{\infty} \left( \tilde \T^i  \X^\top \right) \W.
\end{align}
This follows immediately from \eqref{covariance}. 

We note that although the sum of diffusion matrices $\sum_{i=0}^{\infty} \tilde \T^i$ diverges, equations (\ref{covariance}) and (\ref{hatmatrix}) first evaluate the column differences (via multiplication by $\X^\top$) which causes the sum to converge.

\paragraph{Remark}
Equation \eqref{covariance} can also be written using $\T^\infty$:
\begin{align} \label{tinfty}
\bC = \frac{1}{2} \X \D^{-1} \sum_{i=0}^{\infty} \left(\tilde \T - \T^\infty \right)^i  \X^\top. 
\end{align}
In contrast to $\sum_{i=0}^{\infty} \T^i$ and $\sum_{i=0}^{\infty} \tilde \T^i,$ the geometric series $\sum_{i=0}^{\infty} \left(\tilde \T - \T^\infty \right)^i$ is convergent; for details see Appendix \ref{Tinfty}.

\paragraph{Alternative: Diffusion with an absorbing node} 
As an alternative to the unlimited walk as described above, we can choose any treatment as the baseline, which means reducing the design matrix $\X$ by the respective column, giving $\X_r$, and reducing the other matrices by the respective column and row, giving $\D_r$ and $\T_r$ with dimension $n-1$. We call node $r$ the \emph{reference node}. As described in Appendix \ref{reference}, the reference node becomes absorbing, which means that all walks eventually end up there and the geometric series $\sum_{k=0}^\infty\T_r^k$ becomes convergent. We then have
\begin{align} \label{absorbing}
\bC = \X_r \D_r^{-1} \sum_{i=0}^{\infty} \T_r^i  \X_r^\top
\end{align}
which holds in general, also for bipartite graphs.  

Using a simplified version of the proof in Appendix \ref{proof}, it is possible to show that, for the special case of a non-bipartite graph, the covariance and hat matrices can also be found by replacing the reduced matrices $\X_r$, $\D_r$ and $\T_r$ in equation \eqref{absorbing} with their unreduced counterparts.

\subsubsection{Finding NMA estimates iteratively} \label{iteration}

Equations \eqref{covariance} to \eqref{absorbing} mean that we can obtain the covariance matrix $\bC$ and the hat matrix $\H$ iteratively, without computing an inverse or pseudoinverse matrix. Given a vector of observed relative treatment effects $\y$, we can estimate the true effects by the consistent solution $\hat \y,$ where
$$\hat \y = \H \y = \frac{1}{2} \X \D^{-1} \sum_{i=0}^{\infty} \left(\tilde \T^i \X^\top \right) \W \y.$$
This suggests approximating $\hat \y$ by a series of vectors
$$\hat \y_N = \frac{1}{2} \X \D^{-1} \sum_{i=0}^{N} \left(\tilde \T^i \X^\top \right) \W \y,$$
which perform the sum up to some finite value $N$.
The goodness of the approximation can be measured by the sum of squared differences $||\hat \y_N - \hat \y||^2$, starting with $Q = ||\y - \hat \y||^2$. The examples (see R script) often show a remarkably good approximation already in the first iteration steps. To understand why, we look at the very first step, $N=0$. For $N=0$, we have $\tilde \T^0 = \I$ and
$$\hat \y_0 = \frac{1}{2} \X \D^{-1} \sum_{i=0}^{0} \left(\tilde \T^i \X^\top \right) \W \y = \frac{1}{2} \X \D^{-1} \X^\top  \W \y.$$
Due to the structure of $\X$, in this first step the observations $\y$ are mapped onto a set of $n-1$ consistent relative effects, $\hat\y_0$ (see section 3.1.2 in \cite{Rucke:Schwa:2014}). 
Particularly, entries of $\y$ from different studies that represent the same comparison are mapped to the same value in $\hat \y_0$.
As $N$ increases, these consistent effects move closer to the set of effects that minimize the sum of squared differences.

\section{Interpretation} \label{interpretation}

For any transition matrix $\T$ (also $\tilde \T, \T_r$), each element (e.g., $\T_{ij}$) describes the probability of a transition from node $j$ to node $i$. Element $(i,j)$ of $\T^2$ is given by
$$\T^2_{ij} = \sum_{k=1}^n \T_{ik} \T_{kj}$$
which describes the probability of arriving at node $i$ after two steps when starting in node $j$. An analogous interpretation holds for $\T^3_{ij}, \T^4_{ij}, \dots$ and so on.

\subsection*{Illustration 1: Length of stay} Consider a walker who starts in node A (step 0),  switches to node B in step 1, then comes back to node A in step 2, and leaves A again in step 3. 
Until now, this walker has spent a total of two steps in node A.
Now consider many such walkers on the network, who have a probability $p_i$ of starting in node $i$.
The expected total time a walker spends in node $i$ after $N$ steps is calculated by counting the number of walkers in node $i$ at step $k$, then summing these counts up to a total number of steps $N$, and dividing the result by the total number of walkers. Therefore, the vector of expected times spent in each node is given by
$$\sum_{k=0}^N\T^k  \p,$$
where $\p = (p_1, \dots, p_n)^\top$ is the starting distribution of the walk. For example, setting $\p$ equal to the first unit vector gives the first column of $\sum_{k=0}^N\T^k$ which contains the expected length of stay in each node up to step $N$ for those who start their walk in node $1$. The larger $N$ becomes, the less this time depends on the starting point.
However, unlike the stationary distribution of the walk in equation (\ref{stationarydist}), which describes the \textit{fraction} of time spent in each node (and converges to the weighted degree of the node for large $N$), the \textit{total} time until step $N$ retains the dependence on the starting point.
We note also, that because we are dealing with discrete time, the expected time spent in a node is the same as the expected number of visits to that node. 

\begin{figure}[h!tb]
  \caption{A network with five colored nodes and corresponding drinks. Walkers are spreading through the network in steps, taking a sip of their own drink at every station they visit.}
  \label{drinks}
  \begin{center}
    \includegraphics[width=10cm]{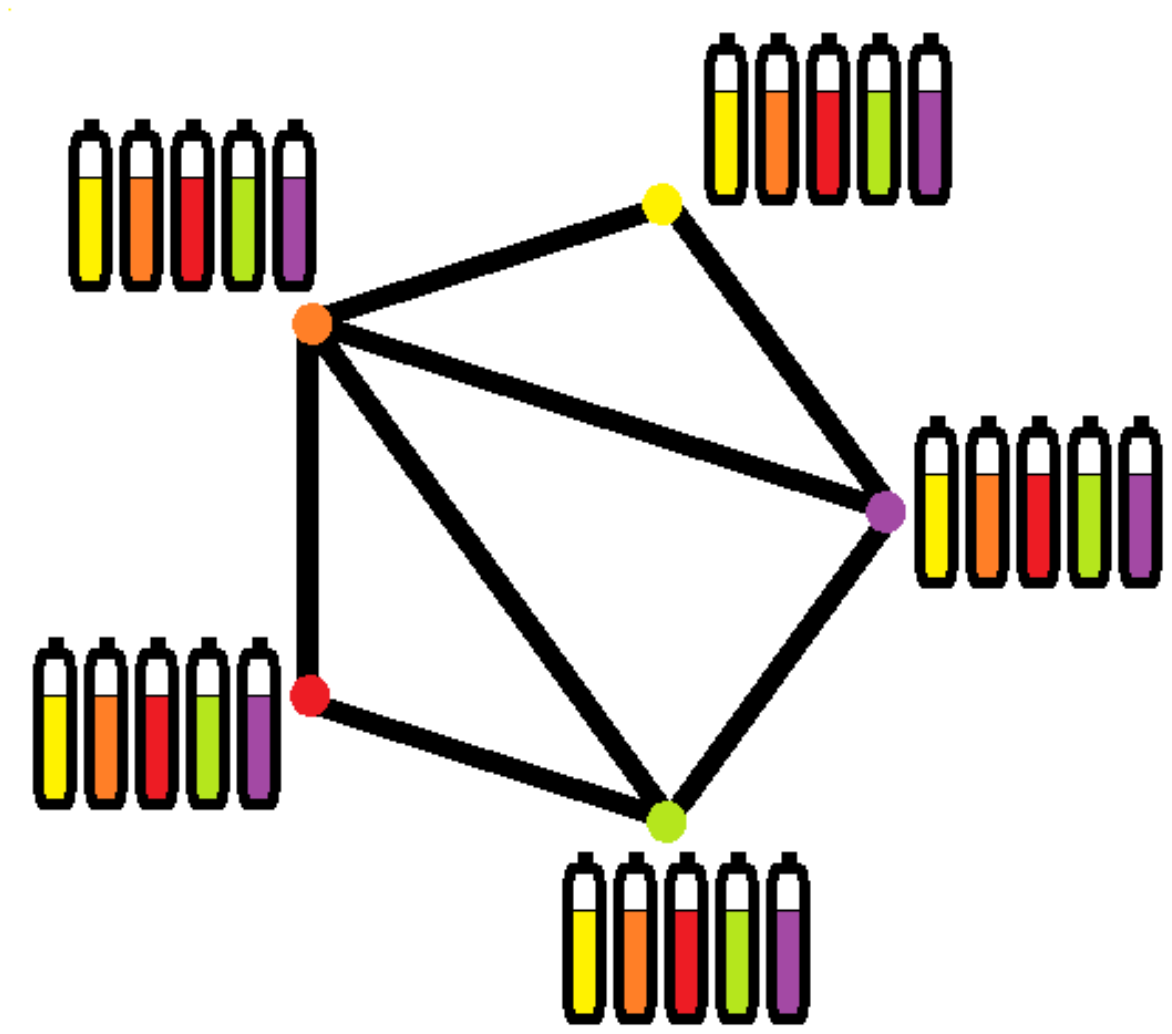}
  \end{center}
\end{figure}

\subsection*{Illustration 2: Walkers and drinks} To provide a perhaps more comprehensible interpretation of times, we now mark the network nodes, for example by placing colored flags such as yellow, orange, red, green, and so on. Large numbers of walkers, wearing T-shirts in the same colors, are placed at ``their own" node. There are also $n$ drinks in various colors available, such as lemon juice, orange juice, red currant juice, kiwi juice, etc. We place $n$ bottles, each with one of the different drinks, at each node, as shown in Figure~\ref{drinks} for Example 1 ($n=5$). As before, we are working with discrete time. Before starting their walk, each walker takes a small, standardized sip of ``their own'' drink at their starting node; for example, ``yellow'' walkers take a sip of lemon juice. Then, all walkers simultaneously start stepwise dispersing through the network, taking each branch with probability proportional to its weight and taking a standardized sip of ``their own'' drink at every station they reach. 
At any time, all walkers have had the same juice intake. However, nodes with a larger (weighted) degree have been visited by more walkers, and the bottles at those nodes contain less juice. 

\paragraph{Scaling - reducing by degree} After a large number of $N$ steps, the remaining juice in each bottle is now further reduced according to the node's (weighted) degree. For example, if the degree is 2, half of every bottle is poured away; if the degree is 3, two thirds of each bottle are removed.  This action, which we call \emph{scaling}, corresponds to multiplying the geometric series $\sum_{k=0}^N\T^k$ by $\D^{-1}$ on the left. 

\paragraph{Column differences} Multiplication with $\X^\top$ on the right means comparing the remaining volume of juice between pairs of bottles of different colors at the same node, for each node in the network. 
These differences are only due to the different starting positions of the walkers. 

\paragraph{Row differences} Multiplication with $\X$ on the left means comparing the differences above between nodes. 
The resulting `differences of differences' in volume then correspond to the covariances in NMA (see below). We note that because we are calculating differences of differences, it is not necessary that the size of bottles at the beginning is equal for each node.
The only requirement is that all bottles must start with enough liquid to facilitate the total number of walkers removing liquid over $N$ steps.

The walkers-and-drinks model suggests an interpretation of the NMA covariance matrix in terms of the long random walk. We measure the difference between the amount of A-juice and B-juice, remaining at A, and the difference between the amount of A-juice and B-juice, remaining at B. We write the difference of differences as
\begin{align*}
(A_A - B_A) - (A_B - B_B) = (A_A + B_B) - (A_B + B_A) = A_A + B_B - 2 \ A_B 
\end{align*}
where we have used the fact that, due to symmetry, $A_B = B_A$ (see Appendix \ref{symmetry}). This may be interpreted as the difference between the fluid intake ``at home'' (i.e., A in A and B in B) and intake ``at one's neighbor's'' (A in B or B in A).
We have implemented the diffusion process in a number of R functions described in the Appendix \ref{R-code}.

\section{Results} \label{results}

\subsection{Example 1} \label{results1}

\begin{figure}[h!tb]
  \caption{Walkers-and-drinks model for Example 1. The bars represent bottles with juice left after 50 steps and scaling. Each block of bars belongs to a node, each bar in a particular color belongs to the walkers from a particular node. Yellow = A,  orange = B, red = C, green = D, violet = E.}
  \label{example1-walk}
  \begin{center}
    \includegraphics[width=12cm]{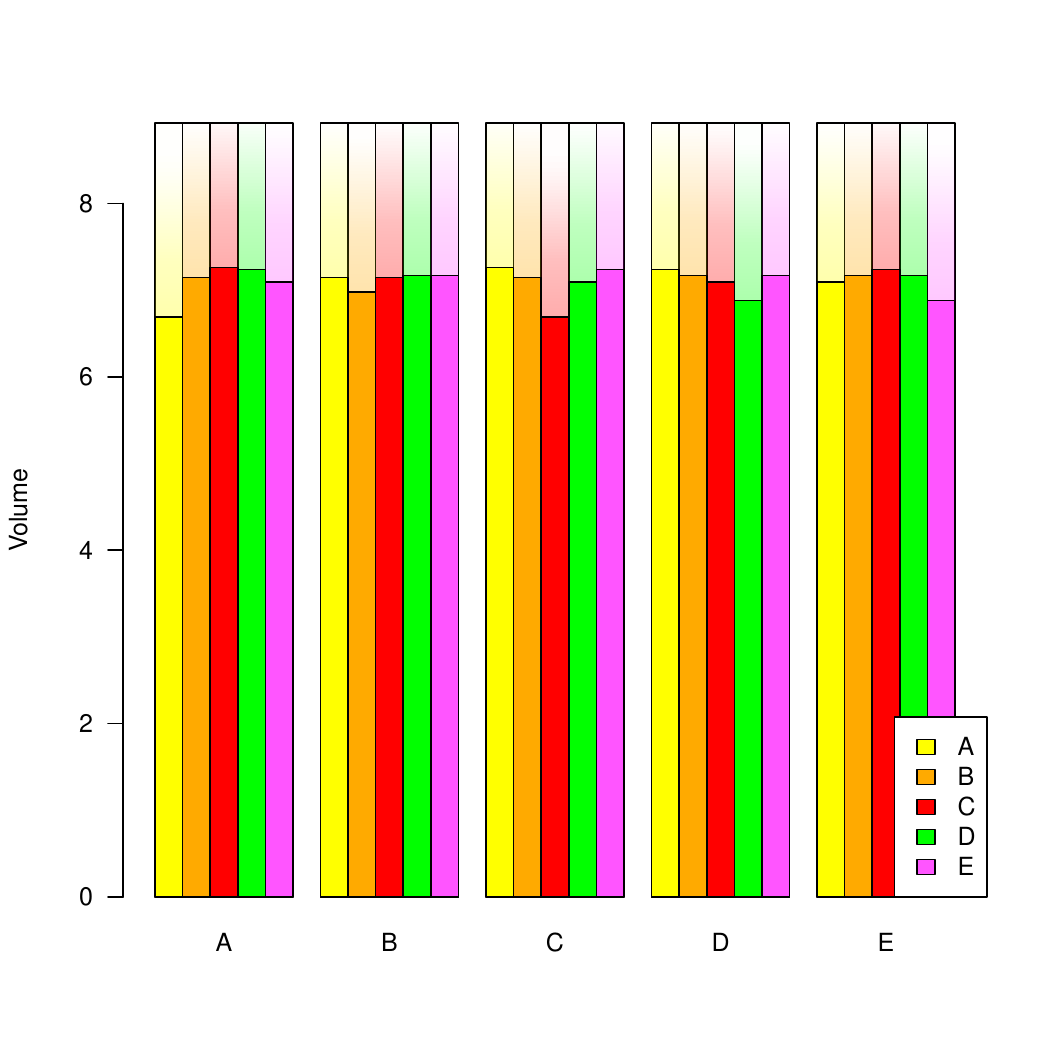}
  \end{center}
\end{figure}

We have shown the diffusion process for the toy example already in Section \ref{sequence} (Figure~\ref{example1-diffusion}). Figure~\ref{example1-walk} illustrates the walkers-and-drinks interpretation. The bars represent the remaining juice in the bottles after 50 walking steps and pouring away part of the juice according to the node's degree. The differences of differences between the nodes correspond to the covariances in NMA. For example, the yellow and the red bar in B have the same height, 7.14, but in A the red bar is higher, 7.26, than the yellow bar, 6.69, with a difference of 0.57. This is the covariance matrix entry for (A:B, A:C).

\subsection{Example 2} \label{results2}

\begin{figure}[h!tb]
  \caption{Iterative estimation of the treatment effects for Example 2. The starting points of the lines (to the left of step 0) are the observed treatment effects from 82 comparisons. The numbers at the bottom are the sum of squared distances of the estimates from the final NMA estimates.}
  \label{example2-draw}
  \begin{center}
    \includegraphics[width=13cm]{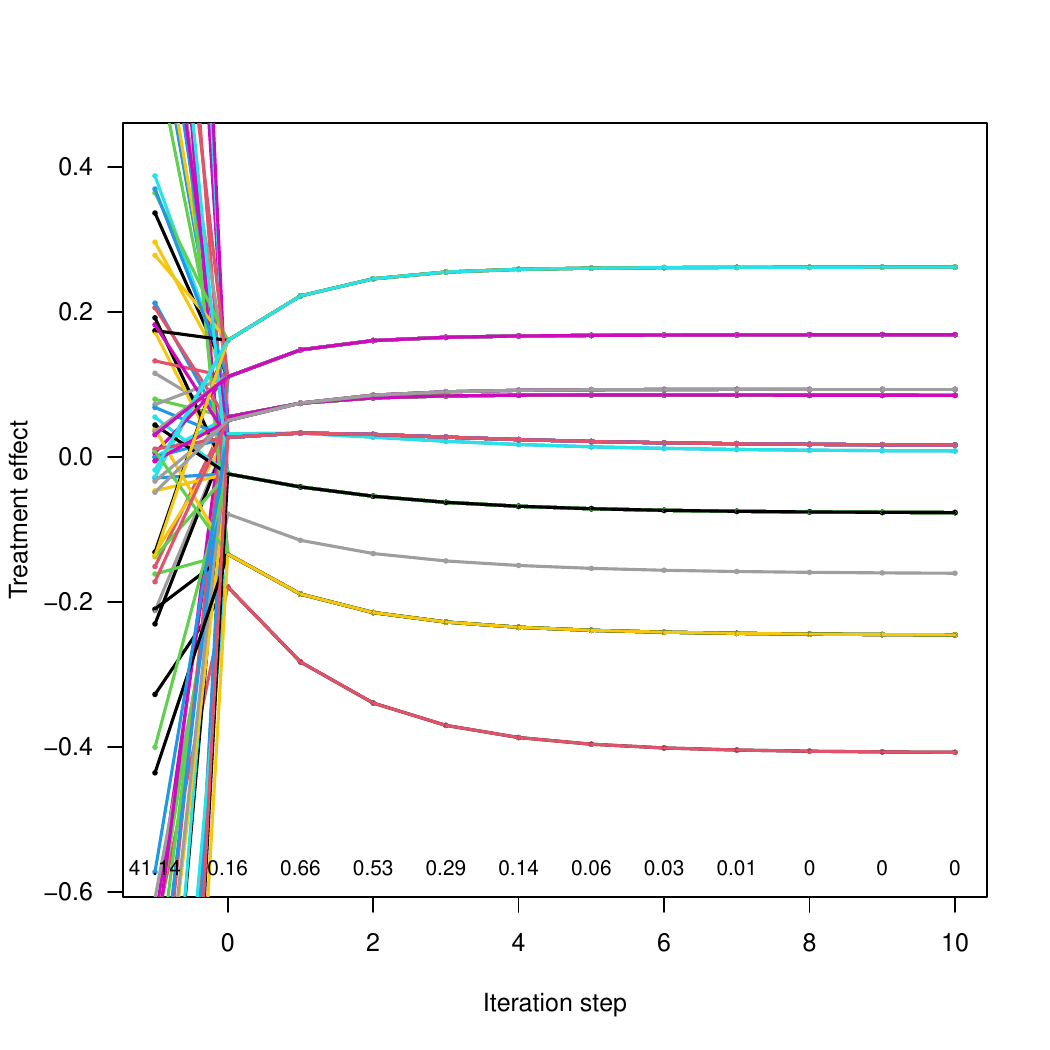}
  \end{center}
\end{figure}

Figure~\ref{example2-draw} shows how the estimated treatment effects of the second example (41 studies with 82 direct comparisons) evolve with each iteration of the geometric series that approximates the hat matrix, here using the lazy walk method. For the comparisons with direct evidence, the treatment effects converge quickly from their direct estimates to their network estimates (note that two pairs of estimates are very similar and cannot be easily distinguished in the figure). As noted in Section \ref{iteration}, the 82 direct observations are mapped to only ten consistent estimates already in iteration step 0.

\begin{figure}[h!tb]
  \caption{Walkers-and-drinks model for Example 2. The bars represent bottles with juice left after 50 steps and scaling. Each block of bars belongs to a node, each bar in a particular color belongs to the walkers from a particular node. Yellow = ICS,  orange = LABA, red = LABA-ICS, green = Placebo, violet = TIO-HH, blue = TIO-SMI.}
  \label{example2-walk}
  \begin{center}
    \includegraphics[width=12cm]{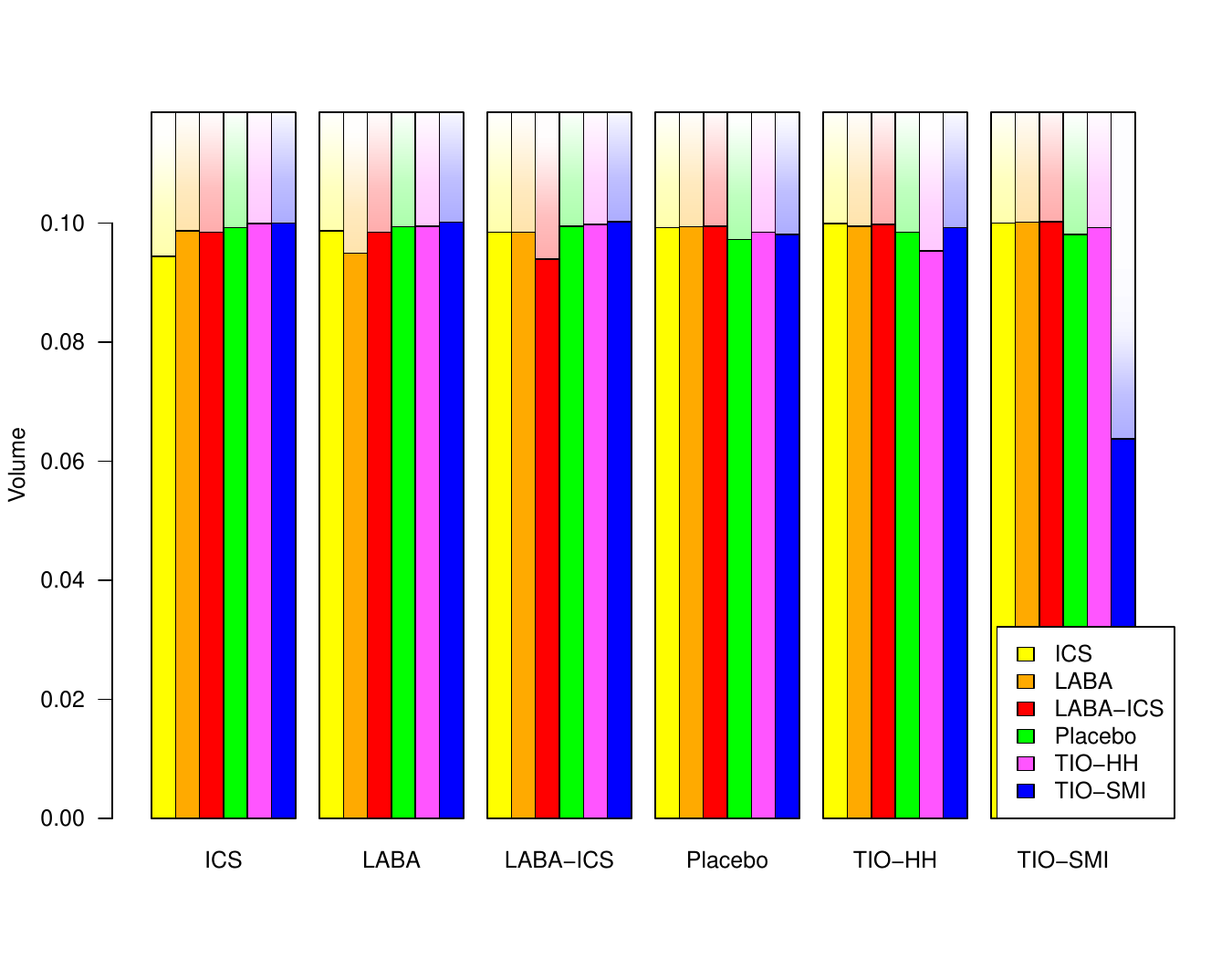}
  \end{center}
\end{figure}

Figure~\ref{example2-walk} illustrates the walkers-and-drinks interpretation for this data. The most noticeable feature of this graph is the short blue bar at its corresponding node, TIO-SMI. As we see from the network graph (Figure~\ref{netgraphs}, middle panel), this node is a spur, connected only to placebo and weakly, via only two trials. The height difference between the blue bar (TIO-SMI) and every other is small at all other nodes but is large at node TIO-SMI. This causes the difference in differences to be particularly large for all comparisons involving TIO-SMI. This makes sense as, by equation \eqref{covariance}, these differences in differences equal the variances of these comparisons, which will be large because they rely heavily on the two studies comparing TIO-SMI and placebo. 

\subsection{Example 3} \label{results3}

\begin{figure}[ptbh]
  \caption{Simple diffusion for Example 3, starting with uniform distribution. The numbers are the percentage weighted degrees.}
  \label{example3-diffusion}
  \begin{center}
    \includegraphics[width=16cm]{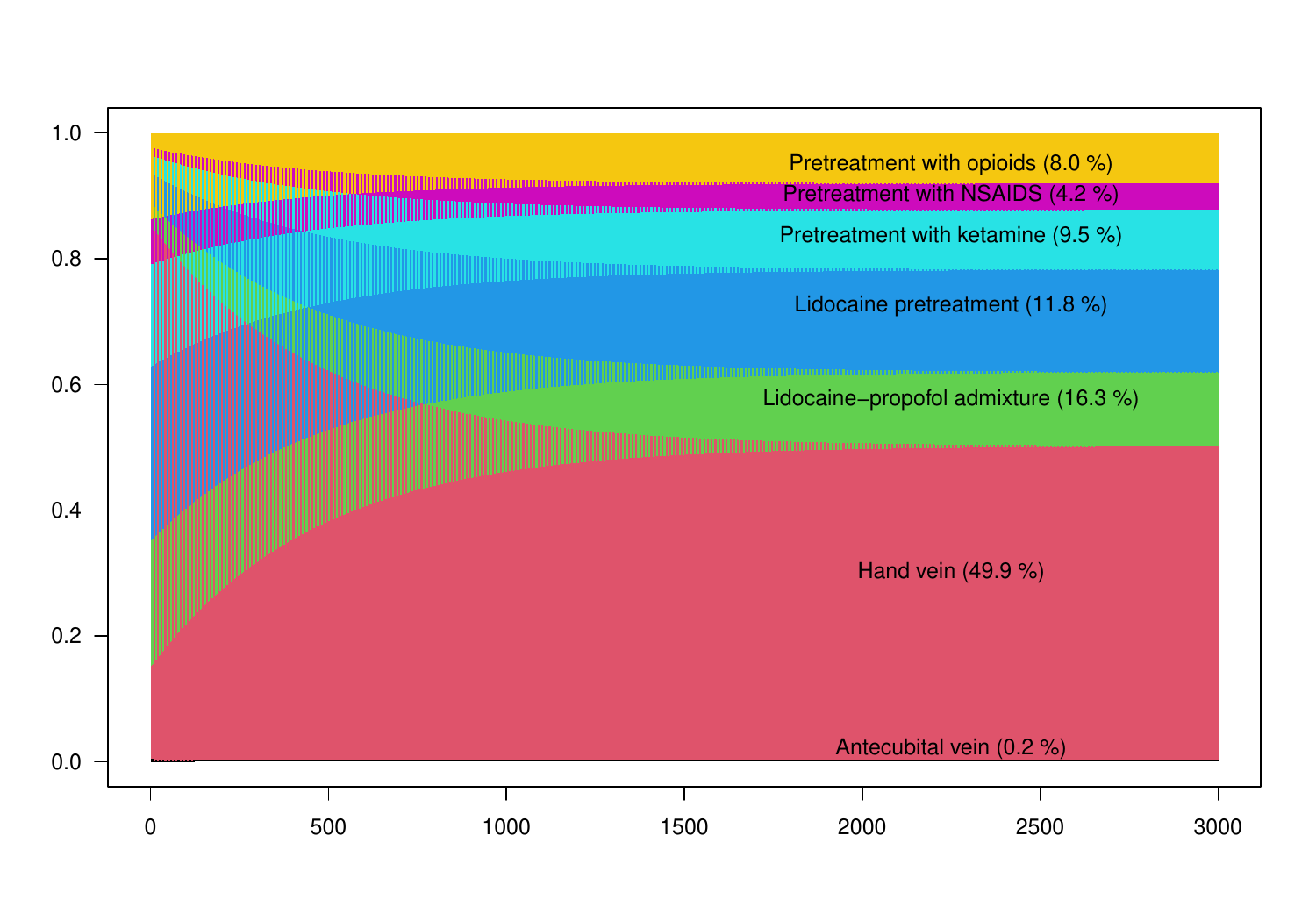}
  \end{center}
\end{figure}

As Figure~\ref{netgraphs} (right panel) shows, the network graph is almost star-shaped, with weak connections between treatments Ante and Lido-prop (1 three-arm study) and between Lido-pre and Lido-prop (2 three-arm studies). This graph is not bipartite, but could be seen as ``almost bipartite'', as demonstrated by Figure \ref{example3-diffusion} which shows 3000 steps of simple diffusion in the form of a stacked probability plot, starting with equal mass at each node and ending with the proportional weighted degrees. We see oscillatory behavior for about 2000 steps.

\begin{figure}[ptb]
  \caption{Iterative estimation of the treatment effects for Example 3. The starting points of the lines are the observed treatment effects from 108 comparisons. Top left: simple method, 1000 steps; Top right: Lazy walk, 10 steps. Bottom left: Absorbing walk, reference is Lidocaine-propofol admixture, 10 steps. Bottom right: Absorbing walk, reference is Hand vein, 10 steps. Numbers at the bottom are the sum of squared distances of the estimates from the final NMA estimates.}
  \label{example3-draw}
  \begin{center}
    \includegraphics[width=16cm]{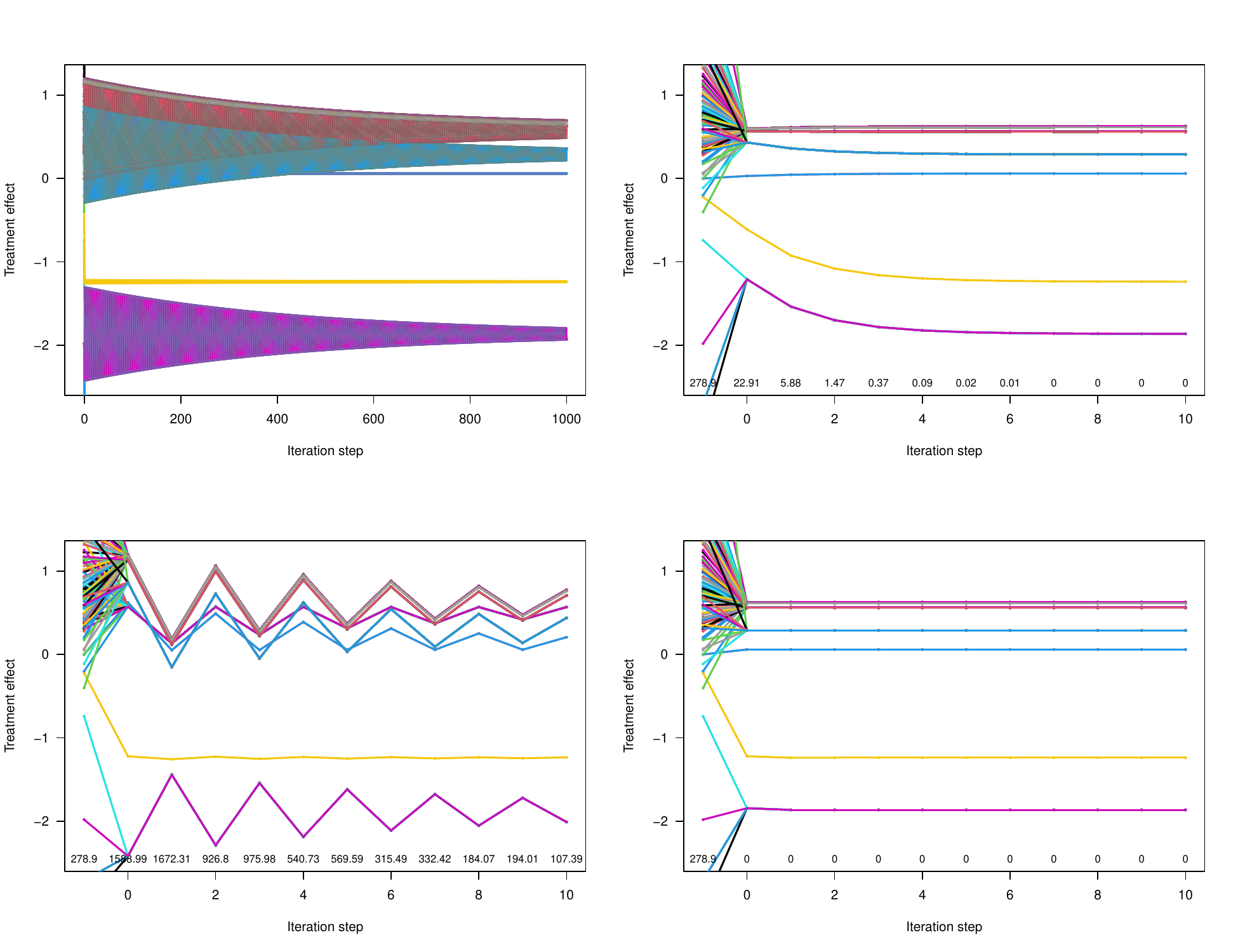}
  \end{center}
\end{figure}

Figure~\ref{example3-draw} shows four versions of the iteration for the relative treatment effect estimates. In Figure~\ref{example3-draw} (top left), we see that the simple walk (equation \eqref{eq:T}) oscillates very strongly, it needs about 3000 steps to converge (see also Figure \ref{example3-diffusion}). By contrast, the lazy walk (equation \eqref{hatmatrix}, top right) converges after only 10 steps. The bottom plots show the convergence of absorbing walks (equation \eqref{absorbing}) with two different reference treatments. The bottom left plot shows oscillatory behavior for the first 10 steps of the absorbing walk with reference treatment Lidocaine-propofol admixture. The bottom right plot shows fast convergence for the absorbing walk with reference treatment Hand vein. This example shows that oscillation may also occur for non-bipartite graphs, and demonstrates that absorbing walks tend to converge most quickly if a central node (i.e., a node with maximum weighted degree) is chosen as reference.

\section{Discussion} \label{discussion}

In this article, we have shown that the covariance matrix of the network estimates in an NMA can be written as a geometric series of diffusion matrices, framed by the design matrix and its transpose. A similar representation exists for the hat matrix. 

This work was partly motivated by a method called network deconvolution that was introduced in the causal inference literature.\cite{Feizi:Marba:Medar:2013,Lin:Pan:Pan:2024} These authors assume that similarities between variables, or even directed causal effects, can be described by some matrix, such as a correlation matrix. The powers of this matrix, suitably scaled, are summed up to provide a matrix of ``total effects'' called the ``transitive closure'' that can be observed. This process is also known as network convolution. The inverse process of identifying the starting matrix, given its transitive closure, is called network deconvolution. In our context of NMA, this would correspond to finding the diffusion matrix from a given covariance matrix, which would be a different task. Apart from that, it seems questionable whether multiplying correlation matrices makes sense, given that correlation is not transitive. 

A geometric series of matrices is also called a Neumann series.\cite{Werner:2018} Finite Neumann series are used in applications to approximate the inverse of a matrix by iteration; for an example, see \cite{Wu:Yin:Voso:Stud:Cava:Dick:2013}.

The iterative calculation of the covariance matrix naturally allows a similar iteration for the hat matrix and thus an iterative method for finding the network estimates that in most of our examples is fast converging. For the version with an absorbing node, the speed of convergence depends on the choice of the reference node. We experienced the best convergence when choosing a central node (for example, placebo as a frequently used control treatment). This is plausible, as a central node rapidly absorbs all the mass. Thus the reduced transition matrix, described in Appendix \ref{reference}, converges quickly to the null matrix, causing the geometric series to attain its limit.

The standard interpretation of a diagonal element of the hat matrix, $\H_{AB,AB}$ in NMA is the proportion of direct evidence from edge AB for comparison A:B.\cite{Dias:Welt:Cald:Ades:chec:2010,Konig:Krahn:Haral:2013,Papak:Nikol:Rucke:2018,Rcker:Papak:Nikol:Schwa:Galla:Davie:2024}
This is also called the leverage: $\H_{AB,AB}$ is the factor by which indirect evidence from the NMA reduces the variance estimate for comparison A:B.\cite{Konig:Krahn:Haral:2013,Rucke:Schwa:2014,Rcke:Nikol:Papak:Sala:Riley:Schw:2020} 
It makes no difference whether the data are provided based on studies (with potentially several studies for the same comparison) or aggregated per comparison.

Another interpretation of the hat matrix comes from the area of electrical circuits. Davies and coauthors showed that entry $\H_{AB,CD}$ can be interpreted as the current flowing through edge CD when a battery is attached across A and B.\cite{Davie:Papak:Nikol:2022} Based on an analogy between electrical networks and random walks, this is equal to the expected net number of times a walker who starts a random walk from node A crosses the edge CD before eventually arriving at node B. Our illustration with times adds another interpretation: If we choose a reference node, say B, this node becomes absorbing, all walks eventually end in B, and the expected times in all other nodes become finite (see Appendix \ref{reference}). It turns out that $\H_{AB,CD}$ corresponds to the scaled  difference in the expected total length of stay in A between walkers who start in C and those who start in D, all walkers eventually arriving in B. 

In fact, our work highlights the potential for even further connections between NMA and the study of complex networks in other disciplines.  
For example, the effective resistance of a graph takes the same form as the variance in NMA.\cite{Rcke:netw:2012}
In the random walk literature, there is an established interpretation for effective resistance in terms of `escape probability', that is, the probability that a walker who starts in node A, reaches node B before they return to A.\cite{DoyleSnell:1984}
More recently, Estrada (2025) interpreted effective resistance as a diffusion distance.\cite{Estrada:2025}  
By defining a diffusion process on a graph, he shows that the effective resistance between two nodes is equivalent to a difference in mass concentrations between the nodes, relative to their initial conditions.
This result is closely connected with our own, and warrants further investigation.

Bipartite networks, which occur frequently in NMA, have a special structure that must be taken into account. A bipartite network graph must not be confused with a different application of bipartite graphs that occurs when an NMA is represented as a hypergraph, as suggested recently.\cite{Lumley:2024,Davies:2026} A hypergraph is a generalized graph, where generalized ``hyperedges'' (in the context of NMA, studies) can join more than two vertices (here treatments, thus including multiarm studies). Each NMA hypergraph can be interpreted as a bipartite graph that consists of two sets of vertices, one representing the treatments, the other representing the studies; both sets are joined by edges where a treatment was investigated in a study. 

In the present article, we concentrated on the common effects model. However, all considerations can be extended to the random effects model by choosing an appropriate weight matrix based on one of the proposed estimators for the heterogeneity variance.\cite{Vero:Jack:Bend:Kuss:Lang:Higg:Kanp:Sala:2019,Lang:Higg:Jack:comp:2019}

We are aware that the phenomena described in this paper, interesting though they may be, have no consequences for statistical practice, as all essential NMA calculations are easily conducted by standard software. Nevertheless, we believe that our findings reveal a nice connection between regression analysis for NMA and the concepts of diffusion and random walks in network analysis.

\begin{appendix}

\section*{Appendix}

\section{Proof of Theorem \eqref{covariance} \label{proof}}
Due to $\T = \I - \L \D^{-1}$ we have $\L \D^{-1} = \I - \T$. Further, it holds $\X = \X  \L^+ \L$, see Appendix A.2 in \cite{Rucke:Schwa:2014}. Using these two relations, we start from the right-hand side of \eqref{covariance}:
\begin{align*}
\frac{1}{2} \X \D^{-1} \sum_{k=0}^{\infty} \left( \tilde \T^k  \X^\top \right)
&= \frac{1}{2} \X \ \L^+\ (\I - \T) \sum_{k=0}^{\infty} \left(\tilde \T^k  \X^\top \right) \\
&= \frac{1}{2} \X \ \L^+\ (\I - \T) \sum_{k=0}^{\infty} ((\T + \I)/2)^k  \X^\top \\
&= \frac{1}{2} \X \ \L^+\ \left( \sum_{k=0}^{\infty} ((\T + \I)/2)^k - \sum_{k=0}^{\infty} \T ((\T + \I)/2)^k \right) \X^\top\\
&= \frac{1}{2} \X \ \L^+\ \left( \sum_{k=0}^{\infty} ((\T + \I)/2)^k - \sum_{k=0}^{\infty} (2(\T + \I)/2 - \I)((\T + \I)/2)^k \right) \X^\top\\
&= \frac{1}{2} \X \ \L^+\ \left( \sum_{k=0}^{\infty} ((\T + \I)/2)^k - 2\sum_{k=0}^{\infty} ((\T + \I)/2)^{k+1} + \sum_{k=0}^{\infty} ((\T + \I)/2)^k \right) \X^\top\\
&= \frac{1}{2} \X \ \L^+\ 2 \left(\sum_{k=0}^{\infty} ((\T + \I)/2)^k - \sum_{k=0}^{\infty} ((\T + \I)/2)^{k+1}\right) \X^\top\\
&= \X \ \L^+ \left(\sum_{k=0}^{\infty} ((\T + \I)/2)^k - \sum_{k=1}^{\infty} ((\T + \I)/2)^{k}\right)  \X^\top\\
&= \X \ \L^+\  ((\T + \I)/2)^0 \X^\top
= \X \ \L^+ \X^\top = \bC
\end{align*}
which is what we wanted to prove.

\section{Version with reference node} \label{reference}
The idea also works, it even simplifies, if we choose an arbitrary baseline treatment, the reference node, and accordingly reduce the design matrix by the column that belongs to the reference node. All other matrices are reduced analogously, which means deleting the row and column that correspond to the reference node. This has interesting consequences. Let us denote the reduced versions by the index $r$. First, the Laplacian matrix $\L_r$ becomes regular, such that $\L_r^{-1}$ exists. Secondly, with respect to diffusion, the reference node becomes absorbing: each walk eventually ends up in this node. This also holds for bipartite networks. For $k \to \infty$ we have $\T_r^k \to \0$ and $\sum_{k=0}^\infty \T_r^k$ is a finite matrix. The same holds for the lazy walk diffusion matrix $\tilde \T_r = (\T_r + \I_r)/2$. For the covariance matrix, we have
\begin{align*}
\bC &=  \X_r \L_r^{-1} \X_r^\top \\
&= \X_r (\D_r - \A_r)^{-1} \X_r^\top \\
&= \X_r \left((\I_r - \T_r)\D_r\right)^{-1}  \X_r^\top \\
&= \X_r \D_r^{-1} (\I_r - \T_r)^{-1}  \X_r^\top \\
&= \frac{1}{2} \X_r \D_r^{-1} (\I_r - (\T_r + \I_r)/2)^{-1}  \X_r^\top \\
&= \frac{1}{2} \X_r \D_r^{-1} (\I_r - \tilde \T_r)^{-1}  \X_r^\top \\
&= \frac{1}{2} \X_r \D_r^{-1} \sum_{k=0}^{\infty} \tilde \T_r^k  \X_r^\top .
\end{align*}
It turns out that for the version with a reference node, bipartite graphs do not need special attention: the results for the walk described by $\T_r$ and for the lazy walk $\tilde \T_r$ agree. 

\section{The behavior of $\sum_{i=0}^\infty(\T - \T^\infty)^i$} \label{Tinfty}

We show equation \eqref{tinfty}. For $i > 0$ we have
\begin{align*} 
\left( \T - \T^\infty \right)^i  = \T^i - \T^\infty  
\end{align*}
which can be shown by complete induction.
Because of $\T^\infty \X^\top  = \0$ we may write
\begin{align*} 
\bC &= \frac{1}{2} \X \D^{-1} \sum_{i=0}^{\infty} \left(\tilde \T^i  \X^\top \right)  - \X \D^{-1} \T^\infty \X^\top \\
&= \frac{1}{2} \X \D^{-1} \sum_{i=0}^{\infty}  \left(\tilde \T^i  -  \T^\infty \right) \X^\top  \\
&= \frac{1}{2} \X \D^{-1} \sum_{i=0}^{\infty}  \left(\tilde \T  -  \T^\infty \right) ^i  \X^\top.  
\end{align*}
Although it has been shown for the average length of stay that $\frac{1}{N} \sum_{i=0}^N \T^i \rightarrow \T^\infty,$\cite{Norris:2007} the following simple example shows that series $\sum_{i=0}^N(\T - \T^\infty)^i$ and $\sum_{i=0}^N(\tilde \T - \T^\infty)^i$ generally do not approach the null matrix for $N \to \infty$.

\paragraph{Example} We consider a simple triangle with all standard errors equal to 1. Then
\begin{align*} 
\T = \frac{1}{2}
\begin{pmatrix}
0 & 1 & 1 \\
1 & 0 & 1 \\
1 & 1 & 0 
\end{pmatrix}, \quad
\T ^\infty = \frac{1}{3}
\begin{pmatrix}
1 & 1 & 1 \\
1 & 1 & 1 \\
1 & 1 & 1 
\end{pmatrix}, \quad
\T  - \T ^\infty = \frac{1}{6}
\begin{pmatrix}
-2 & \phantom{-}1 & \phantom{-}1 \\
\phantom{-}1 & -2 & \phantom{-}1 \\
\phantom{-}1 & \phantom{-}1 & -2 
\end{pmatrix}.
\end{align*}
Because $\T  - \T ^\infty$ is regular, we can write
\begin{align*} 
\sum_{i=0}^\infty \left(\T  - \T ^\infty \right)^i =
\left( \I - \left(\T  - \T ^\infty \right)\right)^{-1} 
= \left( \frac{1}{6}
\begin{pmatrix}
\phantom{-}8 & -1 & -1 \\
-1 & \phantom{-}8 & -1 \\
-1 & -1 & \phantom{-}8 
\end{pmatrix}
\right)^{-1} 
= \frac{1}{9}
\begin{pmatrix}
7 & 1 & 1 \\
1 & 7 & 1 \\
1 & 1 & 7
\end{pmatrix}
\end{align*}
which is not the null matrix. A similar consideration holds for $\tilde \T$.

\section{Symmetry} \label{symmetry}

We note that the matrices $\D^{-1} \sum_{i=0}^{N} \T^i$ and $\D^{-1} \sum_{i=0}^{N} \tilde \T^i$ are symmetric. To see this, we need that $\D^{-1} \T$ and $\D^{-1} \tilde \T$ are symmetric:
\begin{align*} 
\D^{-1} \T &= \D^{-1} \A \D^{-1} \\
\D^{-1} \tilde \T &= \D^{-1} (\A \D^{-1} + \I)/2 = (\D^{-1} \A \D^{-1} + \D^{-1})/2
\end{align*}
which follows from the symmetry of $\D$ and $\A$. The same holds for the powers of $\T$ and $\tilde \T$ and their sums and the covariance matrix \eqref{covariance}.

\section{Implementation in R \label{R-code}}

The supporting information to this article contains three R scripts to reproduce the results. Script \texttt{functions.R} contains a number of functions, \texttt{data.R} provides the data sets, and \texttt{analysis.R} runs the functions on the examples. 
We note that \texttt{data.R} and \texttt{analysis.R} provide additional data and analyses that are not shown in this paper.
\begin{enumerate}
\item Function \texttt{Power()} calculates matrix powers. Its arguments are a square matrix $\M$ and a natural number $k$ (the power). The value of the function consists of the $k'$th power of $\M$, $\M^k$, and the partial sum $\sum_{i=0}^k \M^i$ of the geometric series.
\item Function \texttt{diffusion()} can be used to visualize the exchange of mass between the nodes of a network, given a \textbf{netmeta} object, a starting distribution (i.e., a vector) and a power $N$. There are two plotting options, guided by a logical argument \texttt{proportions}. The function provides either a series of network graphs (\texttt{proportions = FALSE}, default) showing how the mass distribution over the nodes develops with $i = 1, \dots, N$, or a colored stacked probability plot to show this in one figure (\texttt{proportions = TRUE}). An argument \texttt{diffusion.type} allows users to choose between a simple walk (\texttt{"simple"}, default), the lazy walk (\texttt{"lazy"}), or a walk with an absorbing node (\texttt{"absorbing"}). Argument \texttt{ref} can be used to specify a reference node.
\item Function \texttt{Hat()} serves to calculate the hat matrix using a matrix power series via equations \eqref{covariance} to \eqref{absorbing}. Its arguments are a \textbf{netmeta} object, a power $N$, and a logical argument \texttt{ref}, \texttt{FALSE} by default. Argument \texttt{ref} can be used to specify a reference node.
\item Function \texttt{draw.TE()} is based on  \texttt{Hat()}. For a given \textbf{netmeta} object and a power $N$, it allows visualizing the iterative course from the given vector of treatment effects \texttt{TE} to the network estimates \texttt{TE.nma.common} and produces a series of essentially decreasing $Q$ values. Users can choose between the various diffusion types (argument \texttt{diffusion.type}) and specify a reference node (argument \texttt{ref}).
\item Function \texttt{draw.hat()}, likewise based on \texttt{Hat()}, helps visualizing the iterative course of the hat matrix diagonal  (argument \texttt{hat}, default is \texttt{TRUE}), or the same for the treatment effect variances. As with function \texttt{draw.TE()}, there is a choice between diffusion types (argument \texttt{diffusion.type}) and a reference node can be specified (argument \texttt{ref}).
\item Function \texttt{walk()} calculates the steps described in the walkers-and-drinks model for a given \textbf{netmeta} object and visualizes them as bar charts.
\end{enumerate}

\end{appendix}



\begin{Backmatter}

\paragraph{Funding Statement}
ALD is funded by the Engineering and Physical Sciences Research Council (EP/Y007905/1). 

\paragraph{Competing Interests}
None.

\paragraph{Data Availability Statement}
All data used in this article are available from the open source environment R. An R script for reproducing the analyses is available from the Supporting Information of this article.


\paragraph{Author Contributions}
GR developed the methods, wrote the first draft, wrote the R functions and revised the manuscript. ALD revised the manuscript. GS revised the manuscript and the R functions. All authors read and approved the final version of the manuscript.


\printbibliography

\end{Backmatter}

\end{document}